\documentclass[11pt]{article}
\usepackage{amsmath, amsthm, amssymb}    
\usepackage{bridges}			
\usepackage{graphicx}			
\usepackage[colorlinks=true, urlcolor=blue, citecolor=black, linkcolor=black]{hyperref}  
\usepackage{subcaption}			

\title{Art Speaks Maths, Maths Speaks Art}

\author{N. Leone\textsuperscript{1} , S. Parisotto\textsuperscript{2} , K. Targonska-Hadzibabic\textsuperscript{3} , S. Bucklow\textsuperscript{4} , 
A. Launaro\textsuperscript{5} ,\\
S. Reynolds\textsuperscript{6}, 
C.-B. Schönlieb\textsuperscript{7} 
\vspace{10pt}\\
University of Cambridge, Cambridge, UK\\
\textsuperscript{1}{\color{blue}{nl343@cam.ac.uk}}, \
\textsuperscript{2}{\color{blue}{sp751@cam.ac.uk}}, \
\textsuperscript{3}{\color{blue}{kt441@cam.ac.uk}}, \
\textsuperscript{4}{\color{blue}{sb10029@cam.ac.uk}}, \\
\textsuperscript{5}{\color{blue}{al506@cam.ac.uk}},\
\textsuperscript{6}{\color{blue}{scr42@cam.ac.uk}}, \
\textsuperscript{7}{\color{blue}{cbs31@cam.ac.uk}} \\
} 

\date{}					

\begin{document}

\maketitle

\thispagestyle{empty}

\begin{abstract}

Our interdisciplinary team \textbf{M}athematics for \textbf{A}pplications in \textbf{C}ultural \textbf{H}eritage (MACH) aims to use mathematical research for the benefit of the arts and humanities. Our ultimate goal is to create user-friendly software toolkits for artists, art conservators and archaeologists. In order for their underlying mathematical engines and functionality to be optimised for the needs of the end users, we pursue an iterative approach based on a continuous communication between the mathematicians and the cultural-heritage members of our team. Our paper illustrates how maths can speak art, but only if first art speaks maths.




\end{abstract}

\section*{Introduction}
\begin{figure}[h!tbp]
	\centering
	\includegraphics[width=4.4in]{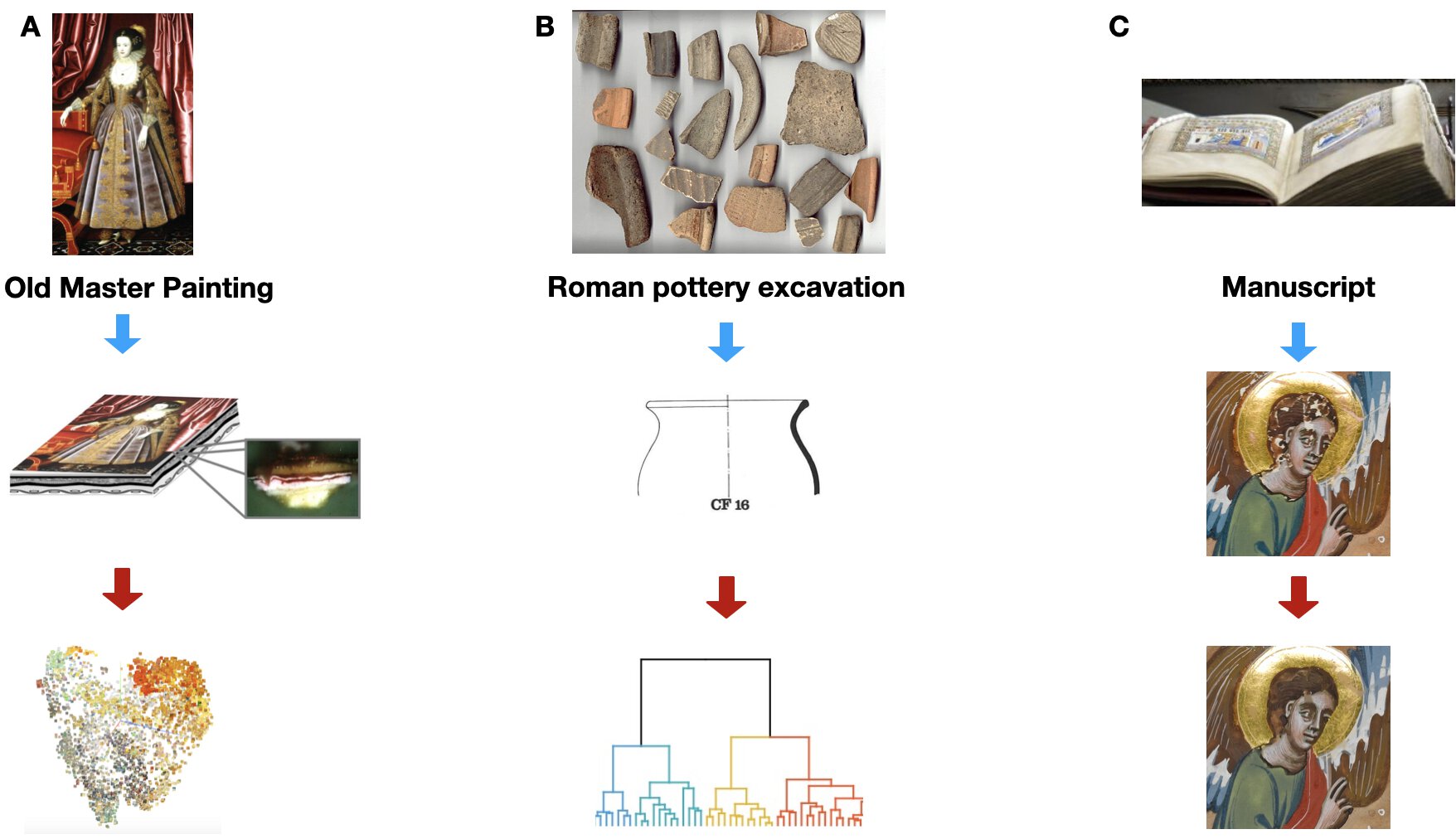}
	\caption{
Scope of the MACH project: (a) Cross-sections of paintings can reveal connections between artefacts from different periods, locales or artists; (b) Mathematical characterisation of pottery shapes can allow their systematic tree-like classification; (c) Digitisation of images of manuscripts allows non-invasive restoration. Emphasis on cross-discipline collaboration is indicated with coloured arrows: "art speaks maths" in blue and "maths speaks art" in red. (Right image: Fitzwilliam Museum, Marlay Cutting It 2, \textcopyright{} Fitzwilliam Museum, Cambridge.)
}
	\label{fig:1}
\end{figure}

Our interdisciplinary work~\cite{Carola} focuses on three specific research areas outlined in Figure \ref{fig:1}. First, we are looking at painting conservation and art history from inside the painting process, by examining the similarities of the images of the cross-sections of paintings, collected in the archives of the Hamilton Kerr Institute in Cambridge, UK. Second, we explore the vast but often inconsistent literature on commonware Roman pottery shapes, and aim to develop their mathematical characterisation in order to provide a coherent and systematic classification.
Third, we investigate the use of advanced mathematical image-processing tools, in particular inpainting, for virtual restoration of illuminated manuscripts, which are rarely restored in their physical form. 

Our goal is to develop mathematical tools which are of true interest to, and effectively usable by the arts and humanities community. In our current work, discussed in this paper, we focus on the projects outlined above, which cover a relatively wide spectrum of challenges in both mathematics and cultural heritage. The successful solution of these challenges requires a continuous communication and feedback between the two communities: the arts and humanities one, which does not have the tools to process the data it already has, and the maths one, which has or could develop such tools, but does not necessarily know what the interesting questions and useful answers are, nor how to communicate the answers in the language most suitable for non-mathematicians. Moreover, the mathematicians in our project are often facing the challenge that there is no objective "ground truth" with which they can compare results and only continuous feedback from the end users can drive useful mathematical developments. Here we illustrate some of the challenges and our approaches to both "art speaking maths" and "maths speaking art", the two aspects necessary in order to get the right answers to the right questions. We are finding that the toolkits that were supposed to be our final products are proving to be indispensable mediators that bridge the two communities within our team.
 
\vspace{-1.em}
\section*{Art Speaks Maths}
The obvious first step is for the arts community to identify interesting questions on which mathematicians could help them. Our collaboration was formed around three such topics, where in each case there already existed data for which mathematical approaches were anticipated to be useful. The next essential step is to pre-process and recast these data in a standardised way that is both faithful (capturing the essence required by the arts community) and suitable for a systematic mathematical analysis. 

In our first project (Figure \ref{fig:1}A) we aim to reveal non-obvious connections between cross-sections (layers of paint) of different paintings, from different artists and/or periods. For example, this could reveal similarities and differences between techniques used to achieve similar effects, such as the ‘optical blues’, the creation of a blue effect by differential scattering rather than absorption. To answer such questions one needs to have, and be able to process, vast amounts of data. We have at our disposal more than 10,000 digital images of cross-sections. The most challenging part of the pre-processing is to isolate the cross-sections from the image backgrounds. This is particularly challenging because the data was collected over 40 years in different ways and archived in different formats. We achieve this using a deep neural network~\cite{Goodfellow}, trained on manually labelled data, and for this training the mathematicians have to rely on verification by the arts experts.
 
In our second project (Figure \ref{fig:1}B), we endeavour to combine and process a broad range of published profiles of Roman commonware pottery (5,000+) into one systematic and coherent classification. Such a comprehensive catalogue will clarify the relationship between different types of pots, in terms of both their chronology and distribution. For example, the spread of certain types in certain periods may point to the development of specific eating habits and/or significant transformations in supply/demand. Here mathematicians need archaeologists not only to provide a clean and consistent data-set of pottery profiles (to be mined from separate published catalogues), but also to validate the resulting classification(s), thus providing the necessary specialist feedback on the results.

In the third project (Figure \ref{fig:1}C), we investigate more than 35,000 digital images from 2000 manuscripts preserved at Fitzwilliam Museum and the Cambridge Colleges. Different manuscripts demand different forms of digital restoration, according to the type of damage: (a) paint losses/lacunae (accidental or intentional), (b) degradation (induced by time or wear-and-tear), (c) abrasion, or (d) overpainting. For example, Duke Francis I of Brittany commissioned overpainting of a lavish prayer book (Fitzwilliam Museum, MS 62) that belonged to his first wife, Yolande of Anjou (1412-1440) with the coat of arms of his new wife, Isabella Stuart (1427-after 1494). The virtual reconstruction is intended to bring back the beauty of the original page, but an art historical perspective is required to judge whether the mathematical reconstruction is historically and stylistically valid.

\vspace{-1.em}
\section*{Maths Speaks Art}
Having understood the requests and requirements from the arts, conservation and humanities communities, and the data pre-processed, the mathematicians can identify the most suitable state-of-the-art methods, develop new ones, and develop or adapt the software for their implementation. 
 
For the cross-sections project we use the technique of content-based image retrieval with k-nearest neighbours. This method is flexible and allows searching for similar patterns in the database, using different metrics and a variety of features to quantify similarity (for example using Gabor filters to identify key textural features, or using moments of the spatial distributions of different colours~\cite{Cross-section}). So far the comparison is done between randomly-selected patches of different pictures. However, more meaningfully for the art conservators, the goal is to segment the images and compare patches belonging to a particular layer of paint.  
 
The classification of pottery profiles is based on the hierarchical clustering of the features extracted from the pottery profiles.
Such features are computed in the latent subspace of sparse autoencoders~\cite{sparse_autoenc}, through their unsupervised training, so as to match the intrinsic sparsity of the shape data-set. An important feature of our approach is that it works well also for the partial profiles obtained from fragmented pottery.
The idea of using hierarchical clustering is based on the necessity to question the classification provided in the catalogues, and check if new clustering patterns can be unveiled. 
Crucially, these objective mathematical results can be correlated with the metadata containing information on, e.g., the supposed type of pottery, the chronology, the location of the excavation, or the manually identified similar profiles, thus allowing an organised inspection of the potentially inconsistent classification in the existing literature.

Virtual image inpainting of illuminated manuscripts first requires the selection of the damaged fragment. 
This can be achieved using a user-supplied inpainting mask or through semi-supervised damage detection via a combination of seed selection and region-growing strategies~\cite{Calatroni}. 
However, different types require different methods. 
For example, tiny elongated scratches in homogeneously-coloured areas are likely to be restored with diffusion-type partial differential equations (see~\cite{Carola_book} for a survey), but large holes in textured regions are visually better reconstructed with \emph{exemplar-based} approaches (see~\cite{exemplar_review} for a survey).
Due to the variety of damages and the numerous methods in the inpainting literature, we collect and test a range of inpainting codes and make them accessible to the conservators and curators at the Fitzwilliam Museum. Furthermore, since a user may want to fuse different details emerging from multi-spectral imaging data, we develop an \emph{ad-hoc} approach based on a variational version of the drift-diffusion osmosis equation~\cite{Simone}, capable of removing the noise of the source channel (e.g.\ the infra-red data), while encoding its relevant structural information into the target visible channel. 
%
\vspace{-1.em}
\section*{Feedback and Toolkits}
In our project, the mathematicians are often challenged by the absence of "the ground truth" - the ideal expected result of their calculations. 
While such ground truth is not known in advance, nor could ever be formulated in a “universal and objective way", the art historians “know it when they see it." 
We therefore established an iterative approach through which mathematicians present their results and the humanities scholars provide feedback and more refined ideas. 

Since the start of the project, our goal has been to develop intuitive software toolkits (as previewed in Figure 2), targeted at the arts-community. 
These toolkits will provide a way for us to give to the world something that the community can use without knowing maths. 
In fact, we are finding that the toolkits, even while they are still being developed, are improving communication and understanding within our team.
They are set up so that users can both use the maths tools intuitively and feed back their own assessment of the quality of the results.

\begin{figure}[htb]
	\centering
	\includegraphics[height=8em]{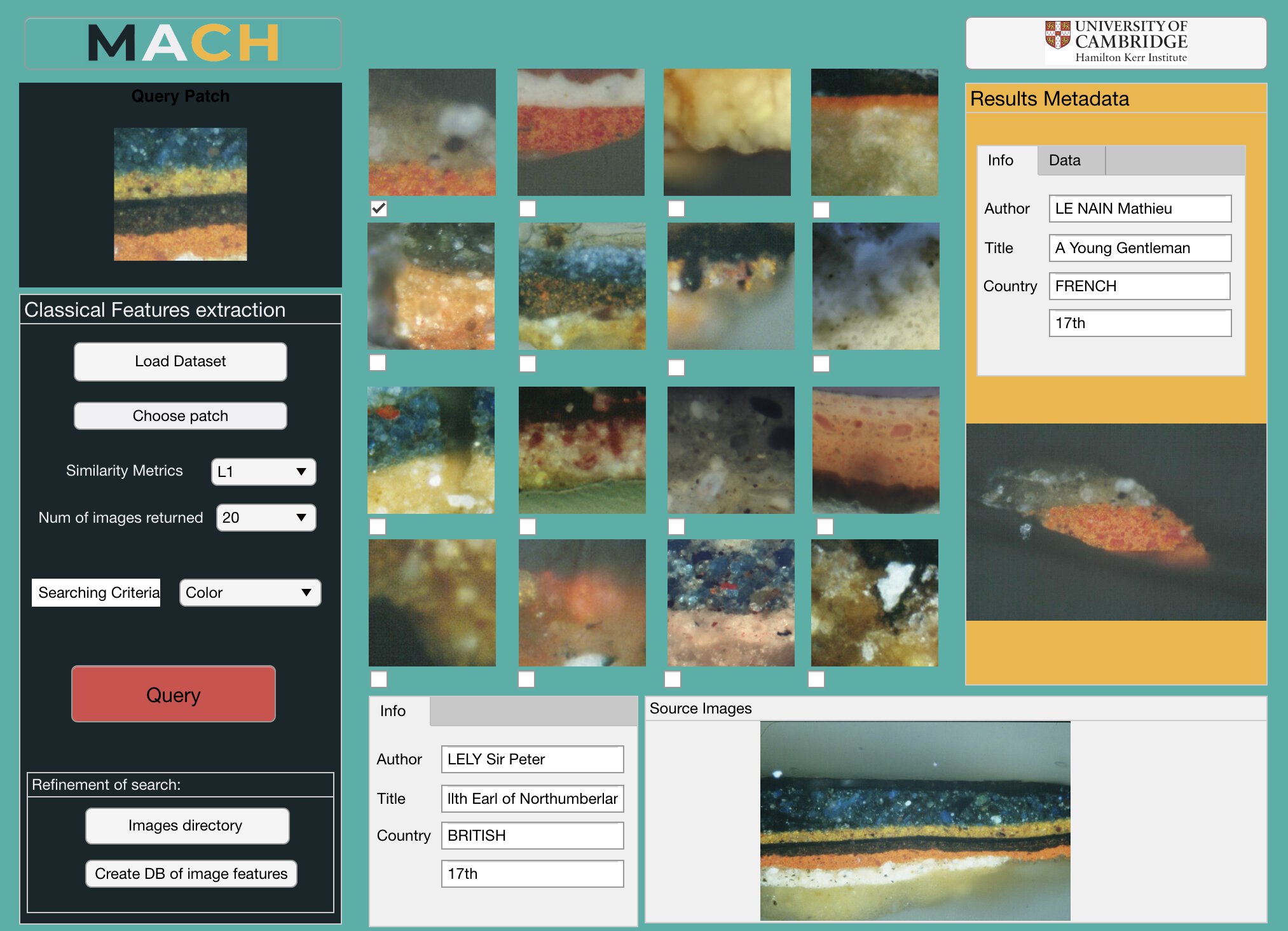}
	\hfill
	\includegraphics[height=8em]{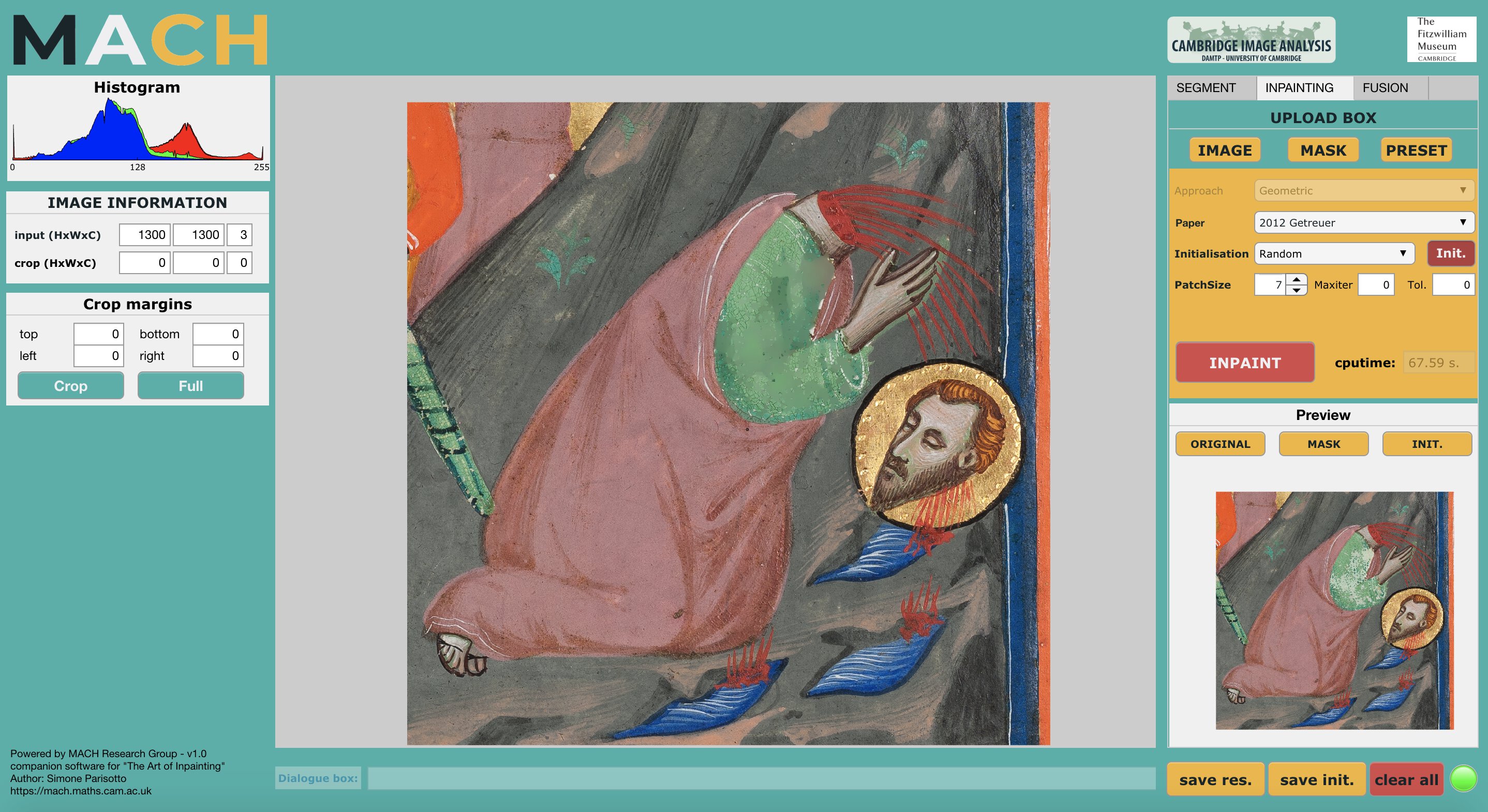}
	\hfill
	\includegraphics[height=8em]{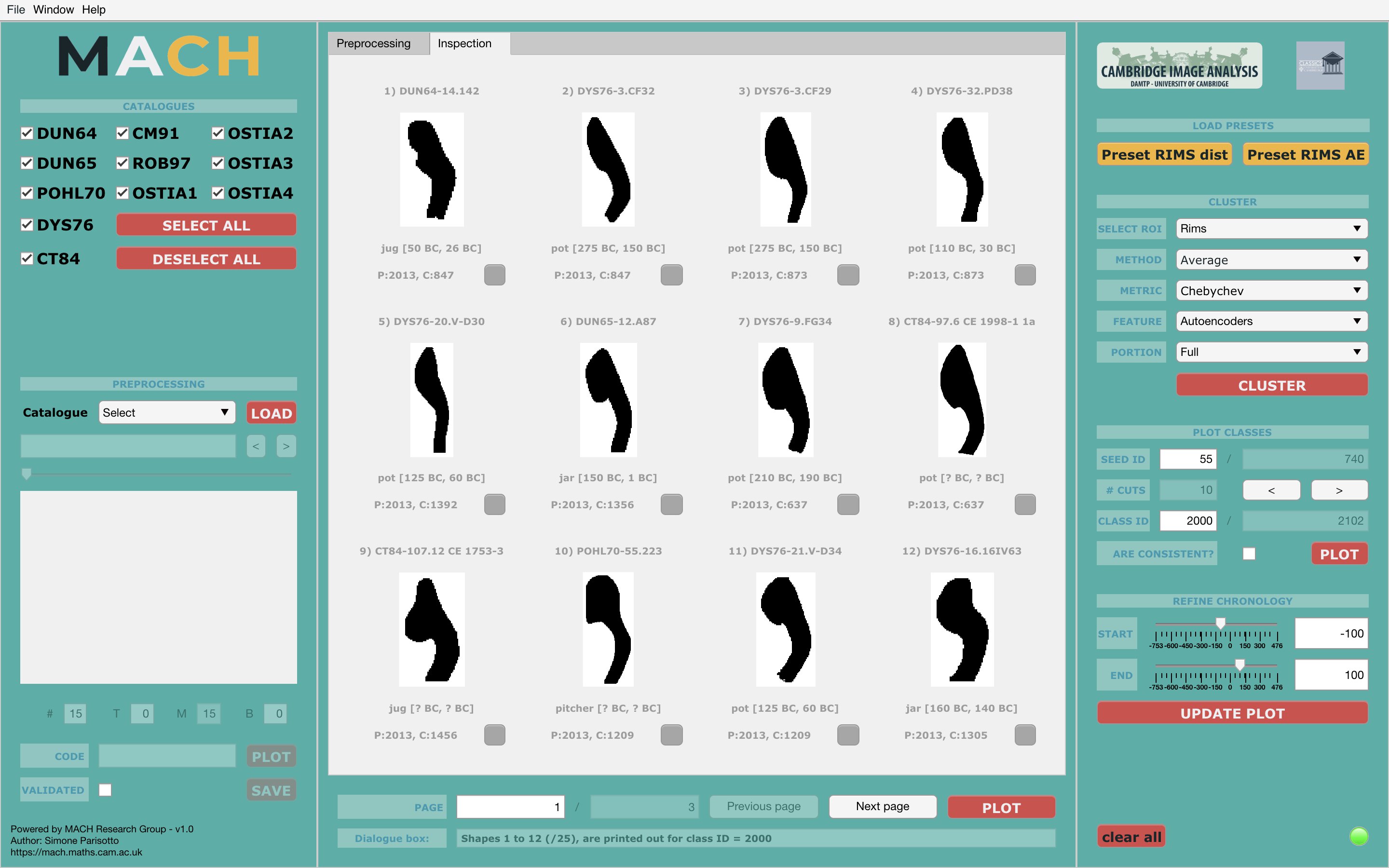}
\caption{Examples of toolkits (under development) for cross-sections analysis of paintings (left), virtual restoration of paintings (middle) and clustering of pottery profiles (right).}
\label{fig:2}
\end{figure}

\vspace{-1em}
\section*{Outlook}
Our first results already show that art can speak maths and vice-versa; in particular, our toolkits provide a favourable platform and conditions for cross-disciplinary interactions. Further improvement of our toolkits will ultimately also require feedback from users in the broader arts and humanities community, and our software will be released on the official webpage \url{http://mach.maths.cam.ac.uk}. We hope that such feedback will continue to challenge and inspire the mathematicians among us, and ultimately lead to new algorithms that artists are not yet even dreaming of.

\vspace{-1em}
\section*{Acknowledgment} The authors acknowledge support from the Leverhulme Trust project Unveiling the Invisible. CBS further acknowledges support from the RISE projects CHiPS and NoMADS, the Cantab Capital Institute for the Mathematics of Information and the Alan Turing Institute.

\raggedright				

\end{document}